\documentclass{article}

\usepackage{amssymb}
\usepackage{graphicx}
\usepackage{dcolumn}
\usepackage{bm}

\begin{document}

\title{Deviation from the Standard Uncertainty Principle
and the Dark Energy Problem}
\author{Shahram Jalalzadeh$^{1}$\thanks{email:
s-jalalzadeh@sbu.ac.ir}, \hspace{.2 cm} Mohammad Ali
Gorji$^2$\thanks{{email: m.gorji@umz.ac.ir}}\hspace{.2cm}
and\hspace{.2cm}Kourosh Nozari$^2$\thanks{email:
knozari@umz.ac.ir}
\\ $^1${\small Department of Physics, Shahid Beheshti
University, G. C. Evin, Tehran 19839, Iran}\\$^2${\small
Department of Physics, Faculty of Basic Sciences,
University of Mazandaran, P. O. Box 47416-95447,
Babolsar, Iran.}}

\maketitle
\begin{abstract}
Quantum fluctuations of a real massless scalar field are studied in
the context of the Generalized Uncertainty Principle (GUP). The
dynamical finite vacuum energy is found in spatially flat
Friedmann-Robertson-Walker (FRW) spacetime which can be identified
as dark energy to explain late time cosmic speed-up. The results
show that a tiny deviation from the standard uncertainty principle
is necessary on cosmological ground. By using the observational data
we have constraint the GUP parameter even more stronger than ever.
\vspace{5mm}
\end{abstract}

\section{Introduction}
The idea that the uncertainty principle is influenced by gravity has
been suggested by many candidates of quantum gravity as well as
string theory. The Generalized Uncertainty Principle (GUP) is an
immediate way to impose the quantum gravity effects in ordinary
quantum mechanics through deforming the usual Heisenberg uncertainty
principle. Such a deformation has origin in the existence of a
minimal measurable length which is predicted by quantum gravity
proposal \cite{QGGUP}. Furthermore, Doubly Special Relativity (DSR)
theories predict an upper bound for the test particles' momentum
\cite{dsr}. Inspired by DSR theories, the UV-regularized version of
GUP has been proposed in \cite{ALI} which supports the existence of
a minimal length and also a maximal momentum. In one dimension, such
deformed uncertainty relation can be written as \cite{nozari1},
\begin{eqnarray}\label{GUP}
\bigtriangleup x\bigtriangleup p \geq \frac{1}{2}\,
\Big(1-\frac{2\alpha_0}{M_{_{Pl}}}\langle p\rangle+
\frac{4\alpha_0^2}{M_{_{Pl}}^2}\langle p^2\rangle\Big),
\end{eqnarray}
where $\alpha_0$ is a numerical factor and $M_{_{Pl}}$ is the Planck
mass \cite{unit1}. The uncertainty relation (\ref{GUP}) predicts the
smallest uncertainty in position $\bigtriangleup
x_{_{min}}=\,2{\alpha_0}\,l_{_{Pl}}$ and a maximum uncertainty in
momentum measurement  $\bigtriangleup
p_{_{max}}=\,M_{_{Pl}}/2{\alpha_0}$. This maximal uncertainty in
momentum measurement gives non-trivially an upper bound also for a
test particle's momentum. The GUP numerical factor $\alpha_0$
defines the quantum gravity scale. But, how much these effects are
small? How one can detect these small corrections? Recently, some
authors attempt to answer these questions. Authors in \cite{DAS}
studied some phenomenological aspects of quantum gravity in quantum
mechanical systems and showed that GUP numerical factor cannot
exceed the Electroweak scale $\alpha_0\leq10^{17}$. In
\cite{nozari2}, the effects of GUP on the transition rate of ultra
cold neutrons in gravitational field have been studied and they
found $\alpha_0\leq10^{29}$, which is weaker than the bound
predicted in \cite{DAS}.

In this paper, we study the effects of the Generalized Uncertainty
Principle (GUP) in cosmology. We show that quantum fluctuations of a
real massless scalar field in Friedmann-Robertson-Walker (FRW)
spacetime, naturally leads to the dynamical UV-regularized vacuum
energy density in GUP framework. We consider the effects of this
vacuum energy density on the expansion rate of the universe and we
find some constraints on the GUP deformation parameter $\alpha_0$.

We stress that while GUP seems to be a UV correction of the standard
uncertainty principle, but as we will show, it is necessary even at
the late time for the renormalizability of the scalar field theory
in a cosmological setup. In fact, as has been stated in Ref.
\cite{Kempf2}, the existence of even an at present unmeasurably
small uncertainty in position (for instance at about the Planck
length) could have a drastic effect in field theory by rendering the
theory to be ultraviolet finite (see also \cite{finite}).

\section{Natural Cutoff}

QFT predicts a divergent vacuum energy for quantum fields. The
common way to resolve this problem is adopting a UV cutoff and
renormalizing vacuum energy to the observed value. On the other
hand, UV cutoff should be determined with quantum gravity theories.
We will see that GUP naturally induces a UV cutoff in QFT.

Taking the vacuum expectation value of the energy-momentum tensor
associated to the real massless scalar field $\phi(x)$, one finds
the well known contribution of the field to the vacuum energy
density,
\begin{eqnarray}\label{nc2}
\rho=\frac{1}{4{\pi^{2}}}\int_0^{\infty}\, dk\,\,{k^3}\,,
\end{eqnarray}
where ${\bf k}$ is the wave vector and $k=|{\bf k}|$. Putting a
cutoff $\Lambda_c$, the vacuum energy density diverges quartically
with cutoff $\rho\sim \Lambda_c^4$. Hence, one has to put a finite
cutoff to get a finite vacuum energy density in QFT.

On the other hand, the modified uncertainty relation (\ref{GUP}) can
be realized from commutation relations \cite{nozari1},
\begin{eqnarray}\label{GUP1}
[\,x_i\,,\,p_j\,]=i\Big(1-\frac{\alpha_0}{M_{_{Pl}}}\,p+
\frac{2\alpha_0^2}{M_{_{Pl}}^2}\,p^2\Big)\delta_{ij}\,,
\end{eqnarray}
\begin{eqnarray}\label{GUP2}
[\,x_i\,,\,x_j\,]=i\frac{\alpha_0}{M_{_{Pl}}}\Big(\frac{4\alpha_0
}{M_{_{Pl}}}-\frac{1}{p}\Big)\Big(p_i x_j-p_j x_i\Big)\,,
\end{eqnarray}
\begin{eqnarray}\label{GUP3}
[\,p_i\,,\,p_j\,]=\,0\,.
\end{eqnarray}
The deformed density of states which is consistent with the deformed
commutation relations (\ref{GUP1}), (\ref{GUP2}) and (\ref{GUP3}) is
obtained in appendix A as
\begin{eqnarray}\label{GUPWF}
\frac{1}{(2\pi)^D}\int_{-\infty}^{+\infty}d^{D}x\,d^{D}p\,
\longrightarrow\,\frac{1}{(2\pi)^D}\int_{_{-\frac{M_{_{Pl}}}
{2\alpha_0}}}^{^{+\frac{M_{_{Pl}}}{2\alpha_0}}}d^{D}x\,d^{D}p
\Big(1-\frac{\alpha_0}{M_{_{Pl}}}\,p+\,\frac{2\alpha_0^2}{
M_{_{Pl}}^2}\,p^2\Big)^{-D},
\end{eqnarray}
where $D$ is the number of degrees of freedom. In appendix A, we show
that the deformed density of states (\ref{GUPWF}) is invariant under
the time evolution and consequently the Liouville theorem is satisfied
in the GUP framework. Neglecting the linear term $-\frac{\alpha_0}{
M_{_{Pl}}}\,p$ and identifying deformation parameters as $\beta=\frac{
2\alpha_0^2}{M_{_{Pl}}^2}$, relation
(\ref{GUPWF}) coincide with the result obtained in Ref. \cite{Liouville}
which supports only the existence of minimal length, not the maximal momentum. Also, neglecting
the quadratic term $\frac{2\alpha_0^2}{M_{_{Pl}}^2}\,p^2$, relation
(\ref{GUPWF}) is in agreement with the result obtained in Ref.
\cite{ALI2} where the author calculated the density of states to
first order of $\alpha_0$. The deformed state density (\ref{GUPWF})
gives the vacuum energy density of the real massless scalar field
$\phi(x)$ in GUP framework as
\begin{eqnarray}\label{nc3}
\rho_{_{bare}}(\alpha_0)=\frac{1}{4{\pi^{2}}}\int_{0}^{\frac{
M_{_{Pl}}}{2\alpha_0}}dk\,k^3\Big(1-\frac{\alpha_0}{M_{_{Pl}}}
\,k+\,\frac{2\alpha_0^2}{M_{_{Pl}}^2}\,k^2\Big)^{-3}=\frac{
\zeta}{\alpha_0^4}\frac{M_{_{Pl}}^4}{16\pi^2},
\end{eqnarray}
where $\zeta=\,\frac{192 \sqrt{7}\arctan(1/\sqrt{7})-77}{1372}
\simeq 0.077674$, is a numerical constant and $|{\bf p}|=
|{\bf k}|=k$ in our units. One can recover relation (\ref{nc2})
in the limit of $\alpha_0\rightarrow\,0$. Clearly, there is a
maximum value for the wave numbers in GUP framework as $k_{max}
={M_{_{Pl}}}/2\alpha_0$, and consequently the integral
automatically converges. An interesting result of this section
is that {\it the vacuum energy density of the scalar field
naturally rendered to be finite in the GUP framework}. The
vacuum energy density (\ref{nc3}) is the bare quantity and can
be renormalized to the observed value by standard renormalization
methods. But in this section, our aim was only to show that this
is a finite quantity in the GUP framework.

\section{Vacuum Energy in FRW Spacetime}

We consider the zero-point quantum fluctuation of a real massless
scalar field $\phi(x)$ in the spatially flat FRW spacetime. The
mode expansion of the field is
\begin{equation}\label{mode}
\phi(x)=\int\frac{d^3{k_c}}{(2\pi)^3\sqrt{2 k_c}}\Big(a_{\bf k}
\phi_k(t)e^{i{\bf k}_c{\bf .x}}+a^{\dag}_{\bf k}\phi^{\ast}_k(t)
e^{-i {\bf k}_c{\bf .x}}\Big),
\end{equation}
where $k_c=\,k\,a(t)$ is the comoving momentum, $a(t)$ is the
scale factor which is the solution of the Friedmann equation and
$\phi_{k}(t)$ is determined by the Klein-Gordon equation in FRW
spacetime,
\begin{eqnarray}\label{K-G11}
\phi_{_k}''+2\frac{a'}{a}\,\phi_{_k}'+k_{c}^2\phi_{_k}=0\,,
\end{eqnarray}
where a prime denotes derivative with respect to the conformal time,
$\eta=\int dt/a(t)$. The energy-momentum tensor for a minimally
coupled real massless scalar field is $T_{\mu\nu}=\partial_{\mu}
\phi\,\partial_{\nu}\phi-\frac{1}{2}\,g_{\mu\nu}\,g^{\sigma\rho}
\,\partial_{\sigma}\phi\,\partial_{\rho}\phi\,$, where $g_{\mu\nu}
=(-1,a^2\delta_{ij})$ is the metric of the spatially flat FRW
spacetime. Taking the vacuum expectation value of the energy-
momentum tensor, one finds the contribution of the scalar field to
the vacuum energy density and pressure \cite{vacummenergy},
\begin{eqnarray}\label{rho}
\rho=\frac{1}{8\pi^2}\int\,d k_{c}k_{c}\Big({|\dot{\phi_{k}}|}^2
+\frac{k_{c}^2}{a^2}{|\phi_{k}|}^2\Big),
\end{eqnarray}
\begin{eqnarray}\label{pressure}
p=\frac{1}{8\pi^2}\int\,d k_{c}k_{c}\Big({|\dot{\phi_{k}}|}^2
-\frac{k_{c}^2}{3a^2}{|\phi_{k}|}^2\Big).
\end{eqnarray}

We note that when the scalar field propagates, the background is
assumed to be fixed. In other words, the field has no effect on the
matter, radiation or the cosmological constant in each epoch, but
the scale factor in each case is different. So it is necessary to
see how the different epochs with different scale factors affect the
propagation of the field. Therefore, we have to compute the vacuum
expectation values (Eqs. (\ref{rho}) and (\ref{pressure})) in all of
the three mentioned cases separately.

\subsection{Radiation domination era (RD)}

The positive frequency solution for the Klein-Gordon equation
(\ref{K-G11}) during RD era is $\phi_{_k}=\,\frac{1}{a(\eta)}\,e^{-i
k_c\,\eta}$ \cite{note}. Plugging this solution into (\ref{rho}),
gives
\begin{eqnarray}\label{RD2}
\rho=\frac{1}{4{\pi^{2}}}\int_{0}^{\infty}\,
dk\,k\,\Big(k^2+\frac{H^2}{2}\Big)\,,
\end{eqnarray}
where $H$ is the Hubble parameter in RD era. The first term in the
right hand side of the above relation is nothing but the flat space
contribution which we have obtained perviously in (\ref{nc2}), and
the second term comes from the curvature of the spacetime.

The energy has no well defined definition in general relativity, but
there is a standard definition for an asymptotically flat spacetime,
the so-called ADM energy \cite{adm}. The ADM definition of the
energy associated to the spacetime with metric $g_{\mu\nu}$ is
$E=\,{\cal{H}}(g_{\mu\nu})\,-\,{\cal{H}}(\eta_{\mu\nu})$, where
${\cal{H}}(g_{\mu\nu})$ is the Hamiltonian of the asymptotically
flat spacetime and $\eta_{\mu\nu}$ is the metric of the flat
spacetime. Therefore, the main idea of the ADM proposal is that,
{\it the energy associated to the flat spacetime, doesn't
gravitate}. Inspired by ADM prescription, one can discard the flat
space contribution from the vacuum energy density (\ref{RD2})
\cite{maggiore}
\begin{eqnarray}\label{ve1}
\rho_{_{bare}}=\,\frac{H^2}{8{\pi^{2}}}\int_0^{\infty}\,dk\,k\,.
\end{eqnarray}
The vacuum energy has its origin only in the curvature of spacetime,
but still it is a divergent quantity. Putting a cutoff $\Lambda_c$,
vacuum energy density diverges quadratically with cutoff
$\rho_{_{bare}}\sim{H^2}\Lambda_{c}^2$. Using the deformed density
of states (\ref{GUPWF}), the vacuum energy density in GUP framework
becomes
\begin{eqnarray}\label{RD4}
\rho_{_{bare}}(\alpha_0)=\frac{H^2}{8{\pi^{2}}}\int_0^{
\frac{M_{_{Pl}}}{2\alpha_0}}dk\,k\Big(1-\frac{\alpha_0}{
M_{_{Pl}}}\,k+\,\frac{2\alpha_0^2}{{M_{_{Pl}}^2}}\,k^2
\Big)^{-3}=\frac{\sigma}{\alpha_0^2} \frac{H^2\,
M_{_{Pl}}^2}{8\pi^2}\,,
\end{eqnarray}
where
$\sigma=\frac{96\sqrt{7}\arctan(1/\sqrt{7})+133}{1372}\simeq\,0.16$
is a numerical constant. Of course, the energy density (\ref{RD4})
is the bare quantity and can be renormalized to observed value by
adding counterterm. But the natural value of the energy density due
to the zero-point fluctuation is given by
\begin{eqnarray}\label{RD5}
\rho_{_{Z}}(\alpha_0)=\epsilon\,\frac{\sigma}{\alpha_0^2}
\frac{H^2\,M_{_{Pl}}^2}{8\pi^2}\,,
\end{eqnarray}
where $\epsilon$ is the renormalization numerical factor of the
order of unity. Note that there is no a priori reason for positivity
of the vacuum energy \cite{maggiore}. Nevertheless, here we assume
it to be positive definite. Plugging
$\phi_{_k}=\frac{1}{a(\eta)}e^{-i k_c\,\eta}$ into relation
(\ref{pressure}) gives the pressure
\begin{eqnarray}\label{p1}
p=\frac{1}{4{\pi^{2}}}\int_{0}^{\infty}\,
dk\,k\,\Big(\frac{k^2}{3}+\frac{H^2}{2}\Big)\,,
\end{eqnarray}
the first term is the flat space result and the second term is the
correction due to the curvature of spacetime. Discarding flat space
contribution and using the deformed density of states (\ref{GUPWF}),
the pressure becomes
\begin{eqnarray}\label{p2}
p_{_{bare}}(\alpha_0)=\frac{H^2}{8{\pi^{2}}}\int_0^{\frac{
M_{_{Pl}}}{2\alpha_0}}dk\,k\Big(1-\frac{\alpha_0}{M_{_{Pl}}}
\,k+\,\frac{2\alpha_0^2}{{M_{_{Pl}}^2}}\,k^2\Big)^{-3}\,=
\frac{\sigma}{\alpha_0^2} \frac{H^2\,M_{_{Pl}}^2}{8\pi^2}\,.
\end{eqnarray}
The bare vacuum energy density and pressure satisfy
$p_{_{bare}}(\alpha_0)=\rho_{_{bare}}(\alpha_0)$, but the
counterterm can be chosen so that the renormalized vacuum energy
density and pressure satisfy $p_{_{Z}}=-\rho_{_{Z}}$, as usually one
assumes \cite{maggiore}. An important issue should be explained at
this stage: we set the equation of state to be $p_Z=-\rho_Z$ because
it is necessary to adopt such an equation of state in order to save
the Lorentz invariance in Minkowski case and also general covariance
of the theory in general. In other words, the equation of state is
not precisely as $p_Z=-\rho_Z$, but we set it to be so in order to
have a generally covariant theory.  This is sufficient in the de
Sitter space since the Hubble parameter is a constant in this space.
However, in other epochs (matter and radiation domination), $H$ is
no longer a constant. In this case, as has been explained after Eq.
(40) of Ref. \cite{maggiore}, one has to consider the total
energy-momentum tensor by incorporation of a possible interaction
between the vacuum energy and other energy-momentum sources. We note
also that once one considers the vacuum as the source of the
energy-momentum, the equation of state parameter is fixed on its
usual $-1$ value. Although for unknown sources the equation of state
parameter is unknown initially and one has to obtain it through
analysis and confrontation with observational data, but here the
source of the energy-momentum is explicitly the vacuum and it is
obvious why we set it to be $-1$.

\subsection{Matter domination era (MD)}

The positive frequency solution of the equation (\ref{K-G11}) during
MD era will be
$\phi_{_k}(\eta)=\,\frac{1}{a}\big(1\,-\frac{i}{k_c\eta}\big)\,e^{-ik_c
\eta}$. Inserting this solution into the (\ref{rho}), gives the
vacuum energy density as
\begin{eqnarray}\label{ds1}
\rho\approx\frac{1}{4{\pi^{2}}}\int_{0}^{\infty}\,
dk\,k\,\Big(k^2+\frac{H^2}{2}\Big)\,.
\end{eqnarray}
Discarding the flat space result, the vacuum energy density in GUP
framework can be obtained through the deformed density of states
(\ref{GUPWF})
\begin{eqnarray}\label{ve2}
\rho_{_{bare}}(\alpha_0)=\frac{H^2}{8{\pi^{2}}}\int_0^{
\frac{M_{_{Pl}}}{2\alpha_0}}dk\,k\Big(1-\frac{\alpha_0}{
M_{_{Pl}}}\,k+\,\frac{2\alpha_0^2}{{M_{_{Pl}}^2}}\,k^2
\Big)^{-3}=\frac{\sigma}{\alpha_0^2}\frac{H^2\,
M_{_{Pl}}^2}{8\pi^2}\,,
\end{eqnarray}
where, again $\sigma\simeq 0.16$. Substituting solution $\phi_{_k}
(\eta)=\,\frac{1}{a}\big(1\,-\frac{i}{k_c\eta}\big)\,e^{-ik_c\eta}$
in relation (\ref{pressure}) and subtracting flat space term, gives
\begin{eqnarray}\label{MDp1}
p_{_{bare}}=\frac{1}{4{\pi^{2}}}\int_{0}^{\infty}\,dk\,
k\,\Big(\frac{H^2}{3}+\frac{9H^4}{32k^2}\Big)\,.
\end{eqnarray}
The first term diverges quadratically with UV cutoff $H^2
\Lambda_c^2$, but the second term diverges logarithmically with UV
cutoff and also requires an IR cutoff. Putting IR cutoff $H$, the
second term produces a term of order $H^4\ln{\Lambda_c}/H$ which is
negligible in the late time \cite{maggiore}. Using the deformed
state density (\ref{GUPWF}), the pressure becomes
\begin{eqnarray}\label{MDp2}
p_{_{bare}}(\alpha_0)=\frac{H^2}{12\pi^2}\int_0^{\frac{
M_{_{Pl}}}{2\alpha_0}}dk\,\Big(1-\frac{\alpha_0}{M_{_{Pl}}}
\,k+\,\frac{2\alpha_0^2}{{M_{_{Pl}}^2}}\,k^2\Big)^{-3}=
\frac{\sigma}{\alpha_0^2} \frac{H^2\,M_{_{Pl}}^2}{12\pi^2}\,.
\end{eqnarray}
In particular one can choose the renormalized vacuum energy density
and pressure so that $p_{_{Z}}=-\rho_{_{Z}}$.

\subsection{de Sitter spacetime}

As pervious sections, substituting positive frequency solution of
the Klein-Gordon equation (\ref{K-G11}) in de Sitter space $\phi_{_k}
(\eta)=\,\frac{1}{a}\big(1\,-\frac{i}{k_c\eta}\big)\,e^{-ik_c\eta}$
into (\ref{rho}), gives the vacuum energy as
\begin{eqnarray}\label{dS1}
\rho=\frac{1}{4{\pi^{2}}}\int_{0}^{\infty}\,
dk\,k\,\Big(k^2+\frac{H^2}{2}\Big)\,,
\end{eqnarray}
where $H$ is the constant Hubble parameter in de Sitter space. The
first term in the right hand side of the above equation is the flat
space contribution which can be eliminated by the ADM prescription.
Using the deformed density of states (\ref{GUPWF}), the bare vacuum
energy density in GUP framework becomes
\begin{eqnarray}\label{dS}
\rho_{_{bare}}(\alpha_0)=\frac{H^2}{8{\pi^{2}}}\int_0^{\frac{
M_{_{Pl}}}{2\alpha_0}}dk\,k\Big(1-\frac{\alpha_0}{M_{_{Pl}}}\,
k+\,\frac{2\alpha_0^2}{{M_{_{Pl}}^2}}\,k^2\Big)^{-3}=\frac{
\sigma}{\alpha_0^2} \frac{H^2\,M_{_{Pl}}^2}{8\pi^2}\,,
\end{eqnarray}
where
$\sigma=\frac{96\sqrt{7}\arctan(1/\sqrt{7})+133}{1372}\simeq\,0.16$
is a numerical constant. Plugging $\phi_{_k}(\eta)=\,\frac{1}{a}
\big(1\,-\frac{i}{k_c\eta}\big)\,e^{-ik_c\eta}$ into the relation
(\ref{pressure}), gives the pressure as follows
\begin{eqnarray}\label{dSp1}
p=\frac{1}{4{\pi^{2}}}\int_{0}^{\infty}\,
dk\,k\,\Big(\frac{k^2}{3}-\frac{H^2}{6}\Big)\,,
\end{eqnarray}
the first term is the flat space result and the second term is the
correction due to the curvature of spacetime. Discarding the flat
space contribution and using the deformed density of states
(\ref{GUPWF}), the pressure becomes
\begin{eqnarray}\label{dSp2}
p_{_{bare}}(\alpha_0)=-\frac{H^2}{24{\pi^{2}}}\int_0^{
\frac{M_{_{Pl}}}{2\alpha_0}}dk\,k\Big(1-\frac{\alpha_0}{
M_{_{Pl}}}\,k+\,\frac{2\alpha_0^2}{{M_{_{Pl}}^2}}\,k^2
\Big)^{-3}\,=-\frac{\sigma}{\alpha_0^2} \frac{H^2\,
M_{_{Pl}}^2}{24\pi^2}\,.
\end{eqnarray}
The bare vacuum energy density and pressure satisfy $p_{_{bare}}
(\alpha_0)=-{1/3}\,\rho_{_{bare}}(\alpha_0)$, but again the
counterterm can be chosen so that the renormalized vacuum energy
density and pressure satisfy the relation $p_{_{Z}}=-\rho_{_{Z}}$ (
see for instance \cite{Akhmedov} and \cite{pady}).

\section{Dark Energy}

The main outcome of the previous sections is that quantum
fluctuations of a real massless scalar field in FRW spacetime lead
to the natural dynamical vacuum energy
\begin{eqnarray}\label{vacuumenergy}
\rho_{_Z}=\epsilon\,\frac{\sigma}{\alpha_0^2}\frac{
H^2(t)\,M_{_{Pl}}^2}{8\pi^2}\,.
\end{eqnarray}

Note that in some sense this relation is similar to the result
obtained in the context of the Holographic dark energy model
\cite{Li,Myung2}. Nevertheless, since we are going to include an
explicit interaction between the vacuum energy and other sources of
the energy-momentum, a deviation from the pure Holographic setup
occurs in our case. Once again, in this interacting scenario to
preserve the general covariance we set  the equation of state
parameter to be $-1$. We note also that there is a UV/IR mixing in
this theory which comes out only because our classical ADM-like
subtraction procedure eliminates the troublesome term diverging in
equations.\\
One can define cosmological parameter $\Omega_{_Z}$ as
\begin{eqnarray}\label{vacuumenergy2}
\Omega_{_Z}=\frac{\rho_{_{Z}}}{\rho_{c}}=\epsilon
\frac{\sigma}{3\pi\alpha_0^2}\,,
\end{eqnarray}
where $\rho_{c}=\frac{3H^2}{8\pi G}$ is the critical energy density
and we have used $M_{_{Pl}}^2=G^{-1}$. One is tempted to identify
this vacuum energy as the dark energy responsible for the late time
cosmic speed-up. But this is not actually the case since time
dependence of the dark energy and cold dark matter (CDM) cannot be
the same from observational grounds (see \cite{maggiore} for more
details). In which follows we model a scenario that contains a
constant contribution of $\rho_{_\Lambda}$ with unknown origin, a
time-dependent dark energy contribution $\rho_{_Z}$ interacting with
dark matter contribution $\rho_{_M}$. This model is usually dubbed
as $\Lambda$ZCDM after \cite{maggiore}. Note that the time
dependence of the dark energy comes just from $\rho_{_Z}$. With
$p_{_Z}=-\rho_{_Z}$, $p_{_\Lambda}=-\rho_{_\Lambda}=-\frac{
\Lambda}{8\pi G}$ and the contribution of the dark energy as
$\rho_{_{DE}}=\rho_{_Z}+\rho_{_{\Lambda}}$, we have
\begin{eqnarray}\label{freidmann}
H^{2}(t)=\frac{8\pi{G}}{3}(\rho_{_M}\,+
\rho_{_Z}+\,\rho_{_{\Lambda}})\,,
\end{eqnarray}
\begin{eqnarray}\label{raychaudhuri}
\dot{H}+H^{2}=-\frac{4\pi G}{3}\rho_{_M}+\frac{
8\pi{G}}{3}\,(\rho_{_{Z}}+\rho_{_\Lambda})\,,
\end{eqnarray}
and the Bianchi identities give the conservation equations as
\begin{eqnarray}\label{conservation}
\dot{\rho}_{_M}\,+\dot{\rho}_{_Z}+\,3H\rho_{_M}=
\,0\,,\qquad\,\dot{\rho}_{_\Lambda}=0\,.
\end{eqnarray}
Using Eq. (\ref{vacuumenergy2}) and combining equations
(\ref{freidmann}) and (\ref{raychaudhuri}) one finds
\begin{eqnarray}\label{thubble}
\dot{H}=\,-\frac{3(1-\Omega_{_Z})}{2}H^2+\frac{\Lambda}{2}\,.
\end{eqnarray}
Integrating the above relation gives the Hubble parameter as
\begin{eqnarray}\label{hubble}
H=\,\bigg(\frac{\Lambda}{3(1-\Omega_{_Z})}\bigg)^{
\frac{1}{2}}\,\frac{1+e^{-\sqrt{3(1-\Omega_{_Z})
\Lambda}t}}{1-e^{-\sqrt{3(1-\Omega_{_Z})\Lambda}t}}\,.
\end{eqnarray}
Note that at the late time the Hubble parameter tends to the de
Sitter constant value $H(t\rightarrow\infty)\simeq\sqrt{\frac{
\Lambda_{eff}}{3}}$ where $\Lambda_{eff}=\frac{\Lambda}{
1-\Omega_{_Z}}$. The scale factor at the late time is
$a(t)\propto\,e^{\sqrt{\frac{\Lambda_{eff}}{3}}t}$.

Using the relations (\ref{thubble}) and (\ref{hubble}), we obtain
the deceleration parameter $q$ as
\begin{eqnarray}
q=-1-\frac{\dot{H}}{H^2}=-1+\frac{6(1-\Omega_{_Z})
\,e^{-\sqrt{3(1-\Omega_{_Z})\Lambda}t}}{\Big(1\,+\,
e^{-\sqrt{3(1-\Omega_{_Z})\Lambda}t}\Big)^2}
\end{eqnarray}
Again, at the late time we attains an accelerating phase of
expansion. The deceleration parameter should be positive for a
universe dominated by matter and therefore
\begin{eqnarray}
\lim_{t\rightarrow 0}q\simeq\frac{2-6\,\Omega_{_Z}}
{4}>0\longrightarrow\Omega_{_Z}<\frac{1}{3}\,,
\end{eqnarray}
which imposes a lower bound on the GUP numerical factor $\alpha_0$
as $\alpha_0>0.2283$, where we have set $\epsilon\sim{\mathcal{O}}
(1)$. This is a strong constraint on the GUP parameter and means
that quantum gravity is inevitable even at large scales and at
late time! From another perspective, while deviation from the
standard prescription is so small, this result shows that
modification of the standard Heisenberg uncertainty principle is
also inevitable.

Integrating the relation (\ref{hubble}) gives the scale factor of
the model

\begin{eqnarray}\label{scale factor}
a(t)=c_1\,e^{-\sqrt{\frac{\Lambda}{3(1-\Omega_{_Z})}}
\,t}\Bigg[2\Big(1-e^{3\sqrt{3(1-\Omega_{_Z})\Lambda}
\,t}\Big)\Bigg]^{\frac{2}{3(1-\Omega_{_Z})}}
\end{eqnarray}
where $c_1$ is the constant of integration. Using the relation
(\ref{vacuumenergy2}) in (\ref{freidmann}) one can find $\rho_{_Z}
=\frac{\Omega_{_Z}}{(1-\Omega_{_Z})}\,(\rho_{_M}+\rho_{_\Lambda})$
which leads to the relation
\begin{eqnarray}\label{rhoz-rhom}
\dot{\rho}_{_Z}=\frac{\Omega_{_Z}}{1-\Omega_{_Z}}
\,\dot{\rho}_{_M}\,,
\end{eqnarray}
plugging this relation into (\ref{conservation}) and integrating gives
\begin{eqnarray}\label{rhom}
\rho_{_M}(z)=\rho_{_M}(0)\,(1+z)^{3(1-\Omega_{_Z})}\,,
\end{eqnarray}
and for the vacuum energy density gives
\begin{eqnarray}\label{rhoz}
\rho_{_Z}(z)=\frac{\Omega_{_Z}}{1-\Omega_{_Z}}
\Big(\rho_{_M}(0)\,(1+z)^{3(1-\Omega_{_Z})}+
\rho_{_{\Lambda}}\Big),
\end{eqnarray}

\begin{eqnarray}\label{rhoDE}
\rho_{_{DE}}(z)=\rho_{_Z}+\rho_{_{\Lambda}}=\frac{
\Omega_{_Z}}{1-\Omega_{_Z}}\Big(\rho_{_M}(0)\,(1+z)^{3(
1-\Omega_{_Z})}+\frac{\rho_{_{\Lambda}}}{\Omega_{_Z}}\Big).
\end{eqnarray}

In table 1, we have shown the bounds on $\Omega_{_Z}$ from different
observational probes and we obtained the corresponding values for
the GUP numerical factor $\alpha_0$. We note that the values of
$\Omega_{_Z}$ used in this table are from Ref. \cite{maggiore} and
we have supposed $\epsilon\sim{\mathcal{O}}(1)$. It is important to
note also that the combined data set CMB+BAO+SNIa best fit for
$\alpha_{0}$ gives a result of order of unity that is far more
stronger than the bounds obtained in \cite{DAS} and \cite{nozari2}.

\begin{table}
\begin{center}
\begin{tabular}{|c|c|c|c|c|c|}
\hline Source & $\Omega_{_Z}$ & $\alpha_0$\\
\hline \textbf{BBN} & $<0.29$ &
 $>0.2448$ \\
\textbf{CMB+BAO+SNIa (best fit)} &
$0.002\pm0.001$
 & $3.28882\pm0.8810$ \\
\textbf{CMB+BAO+SNIa} &
$\leq0.3$
 & $\geq0.2407$ \\
\textbf{time evolution of dark energy} &
$\leq 0.1$
 & $\geq0.4169$  \\
 \hline
\end{tabular}
\end{center}
\caption{\label{tab:table1} Bounds on $\Omega_{_Z}$ from different
sources of cosmological observations and corresponding values for
the GUP numerical factor $\alpha_0$. In all of these results, we
suppose $\epsilon\sim{\mathcal{O}}(1)$.}
\end{table}

\section{Conclusions}
Quantum gravity proposal provides some corrections to the standard
uncertainty principle which is called the Generalized Uncertainty
Principle. There is a free parameter $\alpha_0$ in this theory which
determines the fundamental length of quantum gravity $\alpha_0
l_{_{Pl}}$. It is widely believed that Planck length is the
fundamental length and consequently $\alpha_0$ is of the order of
unity $\alpha_0\sim 1$. In fact, $\alpha_0$ should be fixed via
experiments. Recently, some upper bounds for the $\alpha_0$ has been
obtained in some quantum mechanical phenomena. In this paper, we
proposed a possible relation between $\alpha_0$ and cosmological
observations. We have studied quantum fluctuations of a real
massless scalar field in the spatially flat FRW spacetime within the
GUP framework. We have shown that the vacuum energy density of the
field naturally gets finite value in this framework. The constraint
on GUP numerical parameter obtained in this paper is very tighter
than those obtained in Refs. \cite{DAS,nozari2}. The lower bound for
$\alpha_0$ shows that a very small deviation from uncertainty
principle is necessary on cosmological grounds.\\

{\bf Acknowledgement}\\
We would like to thanks an anonymous referee for his/her very
valuable comments.

\appendix

\renewcommand{\theequation}{A-\arabic{equation}}
\setcounter{equation}{0}
\section{The Deformed Density of States and the Liouville Theorem}
In this Appendix, we consider time evolution of the deformed density
of states (\ref{GUPWF}) and we show that Liouville theorem is
satisfied in the GUP framework. The classical limit of the deformed
commutation relations (\ref{GUP1}), (\ref{GUP2}) and (\ref{GUP3})
can be obtained by replacing the operators with their classical
counterparts and Dirac commutators with Poisson brackets as
$\frac{1}{i}[\,,\,]\rightarrow\{\,,\,\}$. In $D$-dimensions, the
deformed Poisson algebra is given by \cite{nozari1}
\begin{eqnarray}\label{GUP1Class}
\{\,x_i\,,\,p_j\}=\big(1-\alpha\,p+{2\alpha^2}
\,p^2\big)\delta_{ij}\,,\hspace{0.5cm}\nonumber\\
{\{\,x_i\,,\,x_j\,\}}=\alpha\big(4\alpha-\frac{1}
{p}\big)\big(p_i x_j-p_j x_i\big)\,,\nonumber\\
{\{\,p_i\,,\,p_j\,\}}=\,0\,,\hspace{2cm}
\end{eqnarray}
where we have defined $\alpha=\frac{\alpha_0}{M_{_{Pl}}}$. The
deformation to the phase space due to the deformed commutation
relations (\ref{GUP1Class}) can be obtained through a general
transformation in the corresponding phase space which deforms the
phase space volume as \cite{DOS}
\begin{eqnarray}\label{DFRMD-PSV}
\frac{d^{D}x\,d^{D}p}{J},
\end{eqnarray}
where $J(x,p)$ is the Jacobian of the transformation which can be
expressed in terms of the Poisson brackets as \cite{Fityo}
\begin{eqnarray}\label{Jacobian}
J=\frac{1}{2^{^D}D!}\sum_{i_1...i_{2D}=1}^{2D}
\varepsilon_{i_1...i_{2D}}\{J_{i_1},J_{i_2}\}
...\{J_{i_{2D-1}},\,J_{i_{2D}}\},
\end{eqnarray}
where $\varepsilon$ is the Levi-Civita symbol and $J_i$ denotes the
new phase space variables so that for odd $i$ it is a coordinate and
for even $i$ it is a conjugate momentum. In the Jacobian
(\ref{Jacobian}), the coordinate-coordinate Poisson brackets are
always multiplied by the momentum-momentum Poisson brackets. So, the
non-zero coordinate-coordinate Poisson brackets have no contribution
in the Jacobian because the momenta commutes through relation
(\ref{GUP1Class}). Consequently, the Jacobian (\ref{Jacobian})
simplifies to \cite{DOS}
\begin{equation}\label{Jacobian2}
J=\prod_{i=1}^{D}\{x_i\,,p_i\}=\Big(1-\alpha\,
p+{2\alpha^2}\,p^2\Big)^{D}.
\end{equation}
Using the above Jacobian in relation (\ref{DFRMD-PSV}) gives the
deformed phase space volume in the GUP framework
\begin{eqnarray}\label{InvariantWF}
\frac{d^{D}x\,d^{D}p}{\Big(1-\alpha\,p+\,2 \alpha^2\,p^2\Big)^D}\,.
\end{eqnarray}
In the next step we consider the time evolution of the deformed
phase space volume (\ref{InvariantWF}).

The classical equations of motion can be represented with Poisson
brackets in the Hamiltonian formalism as
\begin{eqnarray}\label{EoM}
\dot{x}_i\;=\;-\{x_i,p_j\}\,\frac{\partial {\mathcal{H}}}
{\partial p_j}+\{x_i,x_j\}\,\frac{\partial {\mathcal{H}}}
{\partial x_j}\;,\nonumber\\{\dot{p}_i}=-\{x_j,p_i\}\,
\frac{\partial {\mathcal{H}}}{\partial x_j}\;,\hspace{1.5cm}
\end{eqnarray}
where ${\mathcal{H}}(x,p)$ is the Hamiltonian of the
system. Consider an infinitesimal change for the phase
space variables under time evolution
\begin{eqnarray}\label{Trans}
x_i' & = & x_i + \delta x_i \;, \cr
p_i' & = & p_i + \delta p_i \;,.
\end{eqnarray}
The dynamics of $\delta x_i$ and $\delta p_i$ is given
by relations (\ref{EoM})
\begin{eqnarray}\label{infin-evol}
\delta{x_i}=\{x_i,p_j\}\,\frac{\partial {\mathcal{H}}}
{\partial p_j}\delta{t}+\{x_i,x_j\}\,\frac{\partial
{\mathcal{H}}}{\partial x_j}\delta{t}\;,\nonumber\\
{{\delta}}p_i=-\{x_j,p_i\}\,\frac{\partial {\mathcal{H}}}
{\partial x_j}\delta{t}\;.\hspace{1.5cm}
\end{eqnarray}
An infinitesimal phase space volume evolves through relations (\ref{Trans}) as
\begin{equation}\label{PSVTE}
d^D{x'}\,d^D{p'}=\bigg|\frac{\partial(x'_i,p'_i)}
{\partial(x_i,p_i)}\bigg|\,d^D{x}\,d^D{p}\;.
\end{equation}
The Jacobian can be obtained by using the relations (\ref{Trans})
and (\ref{infin-evol}). Up to the first order of $\delta{t}$, we
have \cite{Liouville}
\begin{eqnarray}
\bigg|\frac{\partial(x'_i,p'_i)}{\partial(x_i,p_i)}\bigg|=
1+\bigg(\frac{\partial}{\partial x_i}\{x_i,x_j\}
-\frac{\partial}{\partial p_i}\{x_j,p_i\}\bigg)\frac{
\partial{\mathcal{H}}}{\partial x_j}\delta{t}\nonumber\\
=1-{\alpha}D\,(4\alpha-\frac{1}{p})\,p_j\frac{\partial{\mathcal{H}}}{\partial x_j}
\,\delta{t}
\end{eqnarray}
where we have used the relations (\ref{GUP1Class}). Substituting the
above relation in the relation (\ref{PSVTE}) gives
\begin{equation}\label{DoPS1}
d^D{x'}\,d^D{p'}=\bigg(1-{\alpha}D\,(4\alpha-\frac{1}
{p})\,p_j\frac{\partial {\mathcal{H}}}{\partial x_j}
\delta{t}\bigg)\,d^{D}x\,d^{D}p\;.
\end{equation}

On the other hand, we should consider the time evolution
of the term on the denominator of the relation
(\ref{InvariantWF}). Using relations  (\ref{EoM}) and (\ref{infin-evol}), to first order of $\delta{t}$, we have
\begin{eqnarray}\label{p2Trans}
{p'}^2=\sum_{i}p'_i\,p'_i=p^2-2(1-\alpha{p}+2{\alpha^2}
{p}^2)\,p_j\frac{\partial{\mathcal{H}}}{\partial x_j}
\,\delta{t},
\end{eqnarray}
and
\begin{eqnarray}\label{pTrans}
p'=\sqrt{{p'}^2}=p-(1-\alpha{p}+2{\alpha^2}{p}^2)\,\frac{p_j}
{p}\frac{\partial{\mathcal{H}}}{\partial x_j}\delta{t},
\end{eqnarray}
where we have used the fact that $\delta{t}$ is small. Using the
relations (\ref{p2Trans}) and (\ref{pTrans}) we have
\begin{eqnarray}
1-\alpha{p'}+2{\alpha^2}{p'}^2=(1-\alpha{p}+2{\alpha^2}{p^2})
{\bigg(1-\alpha(4\alpha-\frac{1}{p})\,p_j
\frac{\partial {\mathcal{H}}}{\partial x_j}\delta{t}\bigg)},
\end{eqnarray}
which gives the result
\begin{eqnarray}\label{DoPS2}
\big(1-\alpha{p'}+2{\alpha^2}{p'}^2\big)^D=\big(1-\alpha{p}
+2{\alpha^2}{p^2}\big)^D\,{\bigg(1-\alpha{D}\,(4\alpha-\frac{
1}{p})\,p_j\frac{\partial H}{\partial x_j}\delta{t}\bigg)}\,.
\end{eqnarray}
Using the relations (\ref{DoPS1}) and (\ref{DoPS2}) we find
\begin{eqnarray}\label{InvariantWF2}
\frac{d^{D}x'\,d^{D}p'}{\Big(1-\alpha\,p'+\,2
\alpha^2\,p'^2\Big)^D}=\frac{d^{D}x\,d^{D}p}{\Big(1-\alpha\,p+\,2
\alpha^2\,p^2\Big)^D}\,
\end{eqnarray}
which ensures that the phase space volume (\ref{InvariantWF}) is
invariant under time evolution and consequently the Liouville
theorem is satisfied in this setup. Also, it is important to note
that the deformed algebra (\ref{GUP1Class}) induces the maximal
momentum (UV cutoff) as
$p_{_{max}}=1/{2\alpha}=\,M_{_{Pl}}/2{\alpha_0}$. So the range of
integrals in the deformed phase space with phase space volume
(\ref{InvariantWF}) should be changed as has been indicated in Ref.
\cite{nozari1}. Taking this results into account and using the
invariant phase space volume (\ref{InvariantWF}), one can obtain the
deformed density of states (\ref{GUPWF}).


\begin{thebibliography}{}
\bibitem{unit1} We work in unit $\hbar=1=c$, where $\hbar$
and $c$ are Planck constant and speed of light in vacuum,
respectively.
\bibitem{note} In fact, one can not always define the
positive and negative frequency modes for the solutions
of the Klein-Gordon equation in curved spacetime.
Nevertherless, in FRW spacetime, it is possible in the
limit of $\eta\rightarrow-\infty$ (see \cite{fulling}).
\end{thebibliography}
\end{document}